\documentclass[prb,a4paper,twocolumn,showpacs]{revtex4-1}	

\usepackage{graphicx}
\usepackage{epsfig}
\usepackage{amsmath}
\usepackage{epstopdf}
\usepackage{color}

\usepackage{hyperref}

\begin{document}

\title{Exchange coupling and Mn valency in GaN doped with Mn and co-doped with Mg}

\author{Mostefa Djermouni}
\affiliation{Laboratoire de Physique Computationnelle des Mat\'{e}riaux (LPCM), Universit\'{e} Djillali Liab\`{e}s de Sidi Bel-Abb\`{e}s, Sidi Bel-Abb\`{e}s 22000 
\& 
Département de physique, Centre Universitaire Ahmed ZABANA de Relizane, Algeria}

\author{Ali Zaoui}
\affiliation{Laboratoire de Physique Computationnelle des Mat\'{e}riaux (LPCM), Universit\'{e} Djillali Liab\`{e}s de Sidi Bel-Abb\`{e}s, Sidi Bel-Abb\`{e}s 22000, Algeria}

\author{Roland Hayn}
\email{roland.hayn@im2np.fr} \affiliation{Aix-Marseille Univ., CNRS, IM2NP-UMR 7334, 13397 Marseille Cedex 20, 
France}

\author{Abdelkader Boukortt}
\affiliation{Universit\'{e} de Mostaganem, Facult\'{e} des Sciences et de la Technologie, ECP3M, Algeria}

\date{\today}

\hspace{1cm}

\begin{abstract} 
We study 1 or 2 neighboring Mn impurities, as well as complexes of 1 Mn and 1 or 2 Mg ions in a 64 
atoms supercell of GaN by means of density functional calculations. Taking into account the electron correlation 
in the local spin density approximation
with explicit correction of the Hubbard term (the LSDA+$U$ method) and full lattice relaxation we determine the
nearest neighbor exchange $J$ for a pair of Mn impurities. We find $J$ to be ferromagnetic
and of the order of about 18 meV in the Hamiltonian $\hat{H}=-2 J \hat{\vec{S}}_1\cdot \hat{\vec{S}}_2$. That
$J$ is only weakly influenced by the $U$ parameter (varying between 2 and 8 eV) and by the lattice
relaxation. From a detailed analysis of the magnetization density distribution we get hints for a ferromagnetic 
super-exchange mechanism. Also the Mn valence was found to be 3+ without any doubt in the 
absence of co-doping with Mg. Co-doping with 
Mg leads to a valence change to 4+ for 1 Mg and to 5+ for 2 Mg. We show that the valence change can already 
be concluded from a careful analysis of the density of states of GaN doped with Mn without any Mg.
\end{abstract}

\maketitle

\section{Introduction}
The quest for a diluted magnetic semiconductor (DMS) having ferromagnetic order with high 
Curie temperature is heavily disputed in the scientific literature. Several candidate materials had been proposed
with contradicting results. Up to now, the highest transition temperature of about 200 K was confirmed in 
GaAs doped with about 10 percent Mn.\cite{Olejnik08}  
It is generally agreed that in that compound the Zener $p$-$d$ exchange
mechanism is at work.\cite{Zener51} 
That mechanism demands the presence of localized magnetic moments and of 
$p$-type hole charge carriers to align the magnetic moments in a ferromagnetic state. In the compound GaAs:Mn 
the Mn$^{2+}$ ion is stable providing at the same time a magnetic moment and a charge carrier. Soon after 
the discovery of ferromagnetism \cite{Matsukura98} in GaAs:Mn, T. Dietl and co-workers proposed 
two other materials having
the prospective of room temperature ferromagnetism: GaN:Mn  and ZnO:Co.\cite{Dietl01} Both materials were 
widely discussed in the scientific literature and we are going to concentrate here on GaN:Mn. 

Unfortunately, the proposal of room temperature ferromagnetism in GaN:Mn could not be generally 
confirmed. Some authors found indeed a high transition temperature close or even higher than room 
temperature, \cite{Reed01,Thaler02,Sasaki02} whereas others found only low Curie temperatures of 
about 10 K for 10 percent Mn doping \cite{Overberg01,Stefanowicz13}
or antiferromagnetic exchange couplings between Mn ions. \cite{Zajac01,Granville10}

Also the theoretical analysis of that material is far from being well advanced and accepted. 
Despite a large number of ab-initio studies devoted to Mn doped GaN, there are still important controversies 
and the values of the exchange couplings are highly disputed.  
Nowadays, the importance of the strong Coulomb repulsion in the incompletely filled $d$-shell of Mn is rather well
accepted. One way of properly taking into account the Coulomb interaction is the introduction of the 
Hubbard $U$ term in the density functional as it is
done by the LSDA+$U$ method. But even within this LSDA+$U$ method the ground state of GaN:Mn is disputed.
If one considers one Mn atom in a supercell with cubic symmetry without breaking this symmetry, the Fermi 
energy $E_F$
lies in a majority spin $d$ band with a rather high density of states (DOS) at $E_F$, i.e. we obtain a 
half-metallic state. \cite{Sandratskii04,Chan08,Nelson15}
But breaking the cubic symmetry and allowing for a Jahn-Teller distortion in combination with a correct treatment
of the Coulomb repulsion in the $d$-shell opens up a 
gap of about 0.8 eV and leads to an insulating behavior. \cite{Stroppa09,Virot11}
The insulating state corresponds to a stable Mn$^{3+}$ state and no hole doping into the 
valence band in difference to the 
situation in GaAs:Mn.
Whether the lattice relaxation also leads to an energy gap for a pair of Mn 
impurities and whether it would significantly influence the magnetic exchange coupling was not clarified up to now.


All the numerous ab-initio studies of exchange coupling between two neighboring Mn atoms in GaN 
point to a ferromagnetic one (see Ref.\ \onlinecite{Bonanni11} for an overview). Nevertheless, there remain several 
problems and inconsistencies. First of all, there is a considerable discrepancy between the 
ab-initio values \cite{Sandratskii04,GonzalezSzwacki11} 
of the order of 10 to 20 meV and
the more analytical ferromagnetic super-exchange calculations \cite{Sawicki12,Blinowski96} 
giving values being 10 times smaller
but leading to the experimentally observed Curie temperatures. On the other hand, an experimental analysis
of the Curie constant \cite{Bonanni11} confirms the rather large ab-initio values of exchange couplings. Also, 
the character of the exchange coupling is disputed\cite{Zhang16} between the ferromagnetic 
super-exchange mechanism
or the Zener double exchange. \cite{Zener51b} And finally, even the Mn$^{3+}$ valency was recently 
questioned in a
theoretical study. \cite{Nelson15}

It is a famous knowledge that Mg doping in GaN leads to hole carriers. The discovery of $p$-type GaN 
was a crucial step to develop 
the now well known white light emitting diodes and was re-compensated by the Nobel prize 
in 2014.\cite{Nobel14} Therefore, 
it is a tempting idea to co-dope GaN with Mn and Mg to combine local magnetic moments with hole-doping
and to eventually increase the ferromagnetic Curie temperature by the 
Zener $p$-$d$ exchange mechanism. 
Unfortunately, that is not what happens. In fact, the doped holes are immediately captured by the 
Mn$^{3+}$ ions changing its valence to Mn$^{4+}$ or even Mn$^{5+}$. That was 
recently established in a 
combined experimental and theoretical study. \cite{Devillers12} 

In view of this large number of inconsistencies we are addressing here several questions
in an ab-initio supercell study of GaN doped with Mn or co-doped with Mn and Mg. 
In extension to previous literature reports we investigate the combined influence of Coulomb correlations in the Mn 
$d$-shell and of lattice relaxations on the magnetic exchange 
coupling $J$ between neighboring Mn ions and on the 
Mn valence in GaN co-doped with Mg. We used the LSDA+$U$ functional as implemented in the 
WIEN2\textit{k} code and 
a 64 atoms supercell with cubic zinc-blende structure. 
The Hubbard $U$-parameter was exclusively applied to 
the incompletely filled d-shell of Mn and varied in between 0 and 10 eV. 

Due to the good accuracy of the all-electron full potential augmented
plane wave code and the proper inclusion of Coulomb interaction and lattice relaxation 
we are able to determine the nearest neighbor exchange coupling $J$ with high precision.
We are going to clarify whether the lattice relaxation which opens a gap for one Mn atom 
in a 64 atom supercell of GaN also leads to a gap for a pair of Mn impurities and whether it would significantly
influence the magnetic exchange couplings. 
From the magnetic moment values, the $d$ shell filling, and from the densities of states we can 
deduce the Mn valency 
in case of co-doping with Mg. We find that one Mg ion in the supercell changes the valence from Mn$^{3+}$ to 
Mn$^{4+}$ and two Mg-ions lead to Mn$^{5+}$.  
As we will show below, the deceiving effect of Mg co-doping can already 
be concluded from an analysis of the DOS of GaN:Mn without any Mg.

\section{Computational methods}



All calculations were done with the Density Functional Theory (DFT) implemented 
in the Wien2\textit{k} code. \cite{Blaha01}
The atoms were represented by the hybrid full-potential (linearized) augmented plane-wave 
plus local orbitals (L/APW+\textit{lo}) method. \cite{Sjostedt00}
A very careful analysis is done to ensure convergence of the total energy in 
terms of the variational cutoff-energy parameter. 
The total energy was determined using a set of 63 $k$-points in the 
irreducible sector of Brillouin zone, equivalent to an $5\times5\times5$ Monkhorst-Pack 
\cite{Monckhorst76}
grid in the supercell. A value of 7 for R$_{MT}$K$_{MAX}$ was used. 
The zinc-blende lattice of GaN (numerically optimized to $a=4.47 $\AA \ for the primitive unit cell) 
is constructed by use of a $2\times2\times2$ 
supercell in primitive lattice structure, resulting in a basis of 64 atoms.
To solve the \textit{Kohn-Sham} equations, the 
exchange-correlation energy, $E_{XC}$, was calculated using the Perdew-Wang LSDA 
\cite{Perdew92}
and the LSDA+$U$ method \cite{Anisimov91}
in the rotationally invariant form of Liechtenstein et al. \cite{Liechtenstein95}
The $U$-parameter of the Mn 3$d$ orbital was used as a free one between 0 and 10 eV. 
But the relevant results depend on $U$ in an unimportant way. Any value between 2 and 8 eV leads to 
similar results.



\section{Manganese doping}

In a first step we investigated {\bf one Mn impurity} in the 64 atom supercell and calculated the gap at Fermi level
for the fully relaxed situation, i.e. including the Jahn-Teller effect. The gap value as a function of the 
Hubbard-$U$ is shown in Table I and the spin-resolved partial DOS projected onto the Mn 3$d$ and 
the 2$p$ orbitals of two different N atoms 
for $U_{eff}=4$ eV in Fig. 1. 

In cubic symmetry, the Mn 3$d$ electrons are split into a 
lower $e_g$ doublet and a higher $t_{2g}$ triplet. The $e_g$ doublet 
is usually merged with the valence band (of mostly N 2$p$ character) and not seen as a 
distinguished peak. On the other
hand, the $t_{2g}$ multiplet peaks are clearly visible in the neighborhood of the Fermi level. 
They have a capacity of three electrons but are 
only filled by two. Without Jahn-Teller effect, in pure cubic symmetry, the $t_{2g}$ levels are 
perfectly degenerate, even within the LSDA+$U$ method. 
\cite{Sandratskii04,Chan08,Nelson15} But the local lattice distortion around the Mn impurity,
i.e. the Jahn-Teller effect, leads to a gap at the Fermi level which is good visible in Fig. 1. 
The filling of the $t_{2g}$ multiplet with two electrons corresponds
to a Mn valency of 3+. The fact that the $t_{2g}$ peaks at Fermi level have a significant 
(30 per cent) $p$-contribution means that 
the Mn$^{3+}$ valent state is in reality a rather extended molecular-orbital like state. Therefore, 
it was recently interpreted as 
a bound state of a local $S=5/2$ Mn state in $d^5$ configuration (Mn$^{2+}$) with a 
surrounding hole having antiparallel spin on the four
neighboring N ligands. \cite{ Nelson15} However, for that interpretation to be valid, there should be a local 
Mn moment
close to 5 $\mu_B$, which is not confirmed by our calculations (see also later). 

It is interesting to note that only the LSDA result ($U_{eff}=0$) is half-metallic but already a small 
Hubbard correction
of $U_{eff}=2$ eV leads to a gap which is stable in a large range of $U_{eff}$ values.  
So, we confirm the result of Refs.\ 
\onlinecite{Stroppa09,Virot11} that the combined influence of Hubbard-$U$ and Jahn-Teller effect 
(lattice relaxation) leads
to insulating behavior. The gap we found is slightly smaller than in Ref.\ \onlinecite{Virot11} but
the general tendency is the same. Calculating the optical data in such a manner corrects the artificial 
Drude-peak behavior
of a seemingly metallic solution. \cite{Boukortt12}

\begin{figure}
\centering%
\includegraphics[width=0.8 \linewidth]{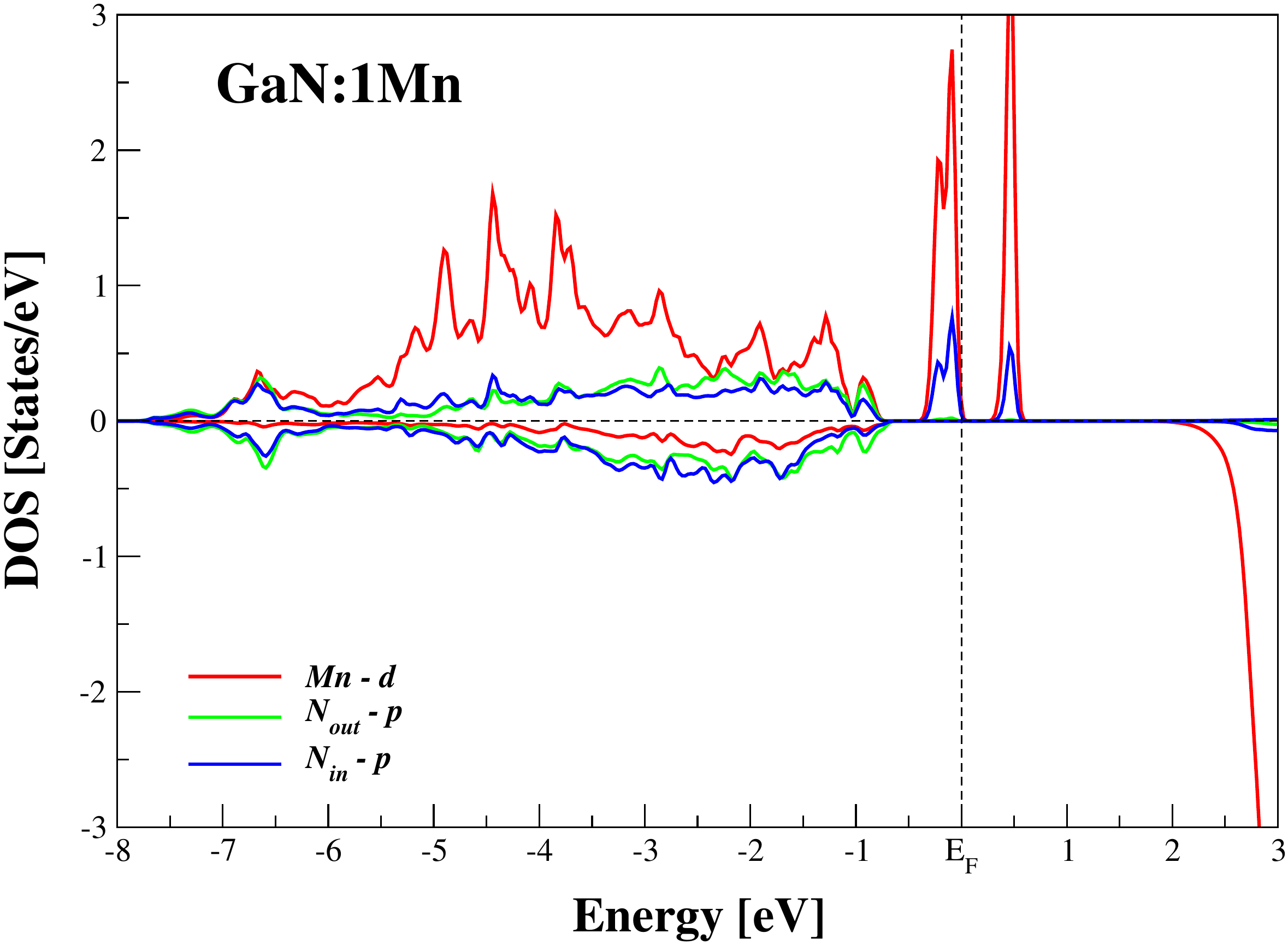}
\caption{ 
(Color online) Projected partial DOS for spin up (upper part) and spin down (lower part) for 
the relaxed structure of one Mn in the 64 atom supercell of GaN. Shown are the partial DOS of the 
Mn 3$d$ orbitals and the 2$p$ orbitals for
one N atom close to the Mn impurity (N$_{in}$) and another one far of it  (N$_{out}$). } 
\end{figure}


\begin{table}
\begin{center}
\begin{tabular}{| c | c | c  |} 
\hline
\hline
$U_{eff}$   & $E_g$ (spin up) & $E_g$ (spin down) \\  
(eV) & (eV) & (eV) \\
\hline
0  &  -- & 2.04 \\
2   &  0.19 & 2.04 \\
4   &  0.42 & 2.09 \\
6   &  0.41 & 2.02 \\
8   &  0.41 & 2.01 \\
10   & 0.30 & 2.00 \\
\hline
\hline
\end{tabular}
\caption{Gap energies of GaN:Mn with one Mn-atom doped in a 64 atom supercell as a function of $U_{eff}=U-J$.} 
\label{label}
\end{center}
\end{table}

\begin{table}
\begin{center}
\begin{tabular}{ | l | l | l | l | l |} 
\hline
$U_{eff}$   & $ \Delta E$ & $M_{total}$ & $M_{Mn} $ & $ M_{N} $ \\
(eV) & (meV) & ($\mu_B$) & ($\mu_B$) &  ($\mu_B$) \\
\hline
0 & 1212.84 & 7.97 & 3.1 & -0.01 \\
\hline
2 & 356.00 & 8.01 & 3.35 & -0.05 \\
\hline
4 & 339.42 & 8.00 & 3.58 & -0.15 \\
\hline
6 & 350.79 & 8.00 & 3.81 & -0.19 \\
\hline
8 & 383.19 & 8.02 & 3.98 & -0.23 \\
\hline
10 & 456.95 & 8.00 & 4.13 & -0.27 \\
\hline
\end{tabular}
\caption{Energy difference $\Delta E = E_{AFM} - E_{FM}$ between AFM and FM arrangements 
of Mn spins for one pair in a 64 atom supercell as a function of $U_{eff}=U-J$. Also given
are total magnetic moment $M_{total}$, as well as the magnetic moments at the Mn-sites
and at the bridging N.}
\label{label}
\end{center}
\end{table}

The super-exchange mechanism between two magnetic ions in an insulating host leads in most of the cases
to an antiferromagnetic exchange coupling. It is therefore highly interesting to check the influence
of the lattice relaxation on {\bf a pair of Mn impurities}. For that purpose, we investigated two 
Mn atoms substituting for two neighboring Ga atoms in the 64 atom supercell (2x2x2 cells of 8 atoms) 
and compared the energy difference between ferromagnetic and antiferromagnetic arrangements
$\Delta E = E_{AFM} - E_{FM}$
as a function of the Hubbard-$U$ parameter (see Table II). 
It can be seen that $\Delta E$ is positive in all cases corresponding to ferromagnetic exchange and it is nearly 
4 times larger in LSDA than in LSDA+$U$. It  varies very slightly with 
$U_{eff}$ once the Hubbard correlation is included. 
We can conclude from Tables I and II that all values of $U_{eff}$ between 2 and 8 eV can be taken as relevant 
ones.

It is very instructive to investigate the influence of lattice relaxation and to look at the total DOS for a pair of Mn 
impurities (Fig. 2 for $U_{eff}=4$ eV
which is also chosen for all the other DOS figures in the present article). 
Without relaxation, the DOS is rather high at 
$E_F$. But a pseudo gap opens due to lattice relaxation leading to a very small DOS at $E_F$. 
On the other hand, the 
lattice relaxation has no essential influence on $\Delta E$, i.e. on 
the exchange coupling. It is
also remarkable that the rather slim $t_{2g}$ double peak for 1 Mn 
impurity broadens considerably into several sub-peaks for a pair of Mn impurities.
  
The energy difference $\Delta E$ can be used to calculate 
the exchange coupling $J$ in the Hamiltonian
\begin{equation}
\label{eq1}
\hat{H}=-2 J \hat{\vec{S}}_1\cdot \hat{\vec{S}}_2
\end{equation}
with the two spin operators $\hat{\vec{S}}_{1/2}$ corresponding to the two Mn-moments. We prefer to use the 
definition of $J$ in (\ref{eq1}) in agreement with the classical works of Larson et al \cite{Larson88} meaning that our 
$J$-values have to be multiplied by two to be compared with the values given in 
Refs.\ \onlinecite{Bonanni11,Sawicki12}. Treating the spin operators as classical vectors of length
$S_{1/2}=S=2$ in agreement with the Mn-valency of +3 in GaN:Mn we obtain for the energy difference
\begin{equation}
\label{eq2}
\Delta E = E_{AFM}-E_{FM} = 4 J S_{class}^2=16 J_{class}
\end{equation}
leading to $J_{class} = (21.2. \ldots 23.8)$ meV for $\Delta E = (340 \ldots 380)$ meV in the representative
region for $U_{eff}$ (2 to 8 eV)

\begin{figure}
\centering%
\includegraphics[width=0.8 \linewidth]{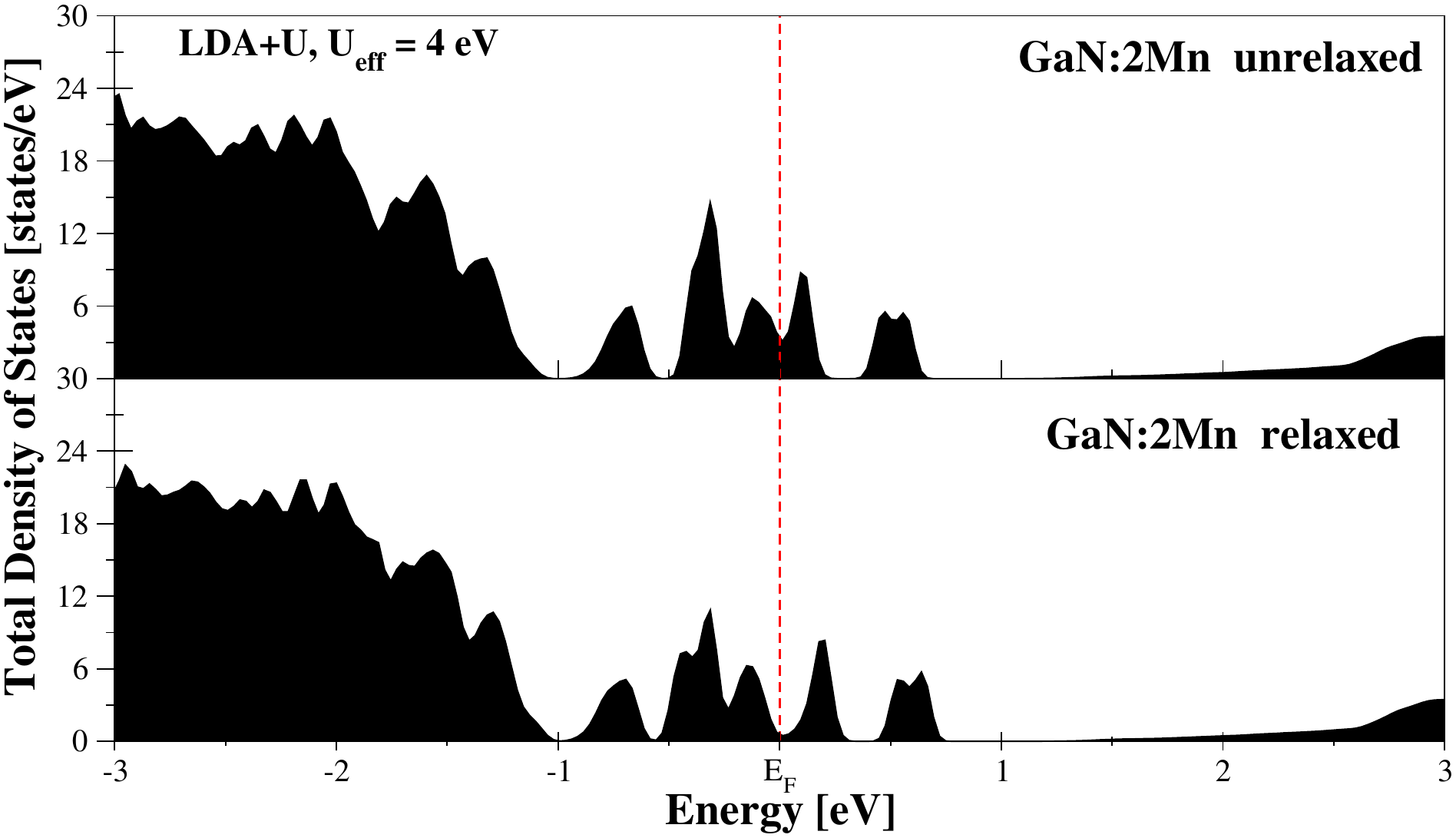}
\caption 
{Total DOS projected onto one spin direction for a ferromagnetic arrangement of one pair
of Mn impurities in the 64 atom supercell of GaN. The atomic positions are not allowed to relax
in the top panel but they are relaxed in the bottom one.}
\end{figure}

It is more justified to treat the spins as quantum operators and $\Delta E$ becomes
\begin{equation}
\label{eq3}
\Delta E = 2S(2S+1) J_{quant}=20 J_{quant}
\end{equation}
with $J_{quant}=(17.0 \ldots 19.0)$ meV.
Our results agree well with former ab-initio values.\cite{Bonanni11}

To provide more detailed information and to get some ideas about the possible exchange mechanism
we analyzed carefully the magnetization density distribution (Fig. 3) and the local magnetic moments on the 
Mn and ligand sites (Fig. 4 left and Table II). It can be observed that the magnetization density is well concentrated
at the Mn sites and the space in between. The local moment at the Mn sites is 3.58 $\mu_B$ far from the 
5 $\mu_B$ of a local $S=5/2$ spin. 
Considerable magnetization contribution anti-parallel to the Mn ones can be found 
on the bridging N (-0.15 $\mu_B$) and smaller ones at the other N ligands. The total magnetic moment
of 4 $\mu_B$ per Mn ion confirms the $S=2$ of a Mn$^{3+}$ state. The good localisation of the magnetization
density on the space in between the two Mn ions in connection with the pseudo gap feature point to a 
ferromagnetic super-exchange mechanism. 

The partial DOS for a pair of Mn impurities in GaN is shown in Fig. 5 (left hand side). It confirms the general picture
for an isolated Mn impurity with the main difference of a significant broadening of the $t_{2g}$ 
complex at the Fermi level.

\begin{figure}
\centering%
\includegraphics[width=0.8\linewidth]{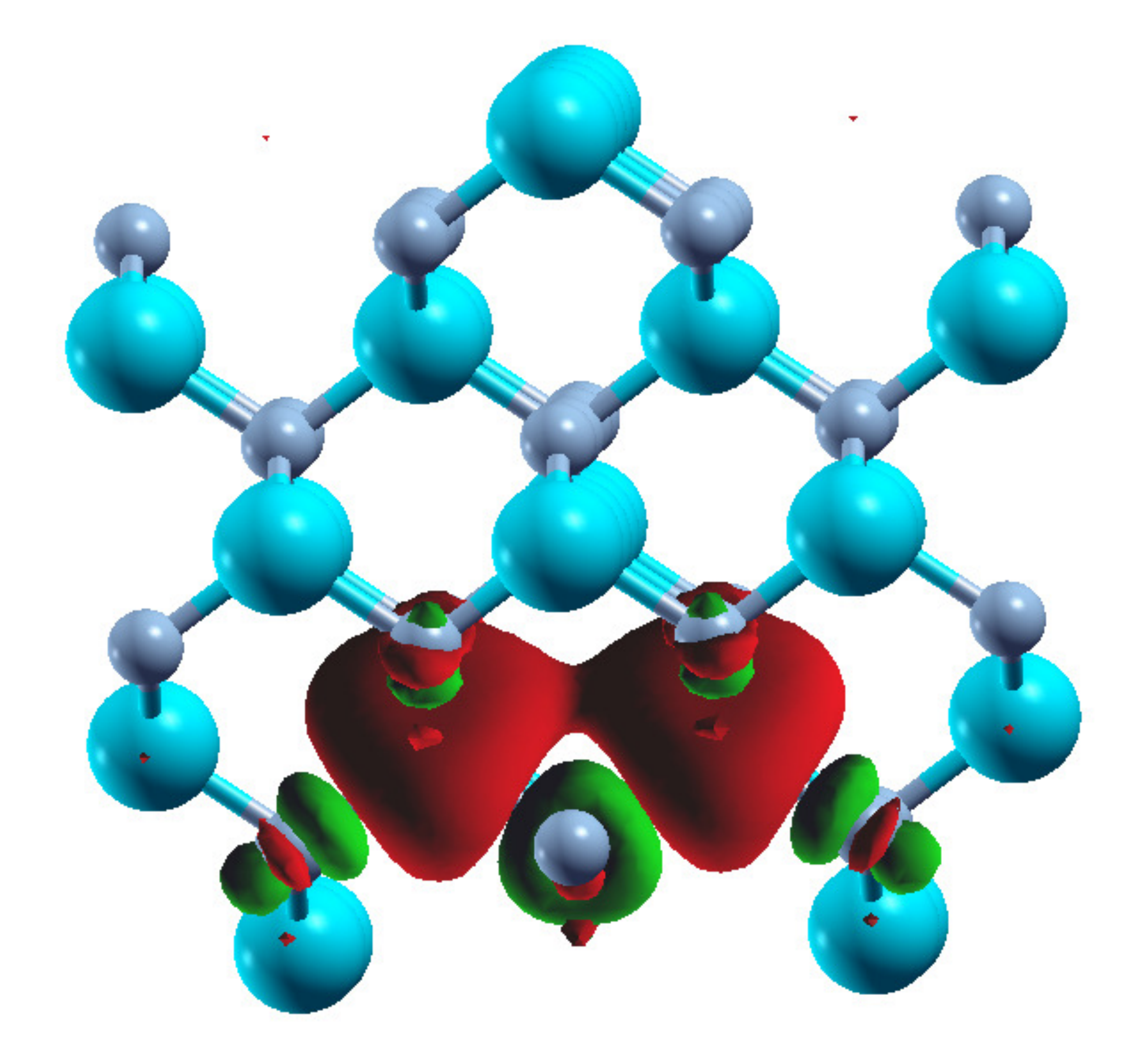}
\caption 
{(Color online) Magnetization distribution for one pair of Mn ions (FM arrangement) in the 64 atom supercell
of GaN.}
\end{figure}

\begin{figure}
\centering%
\includegraphics[width=0.7\linewidth]{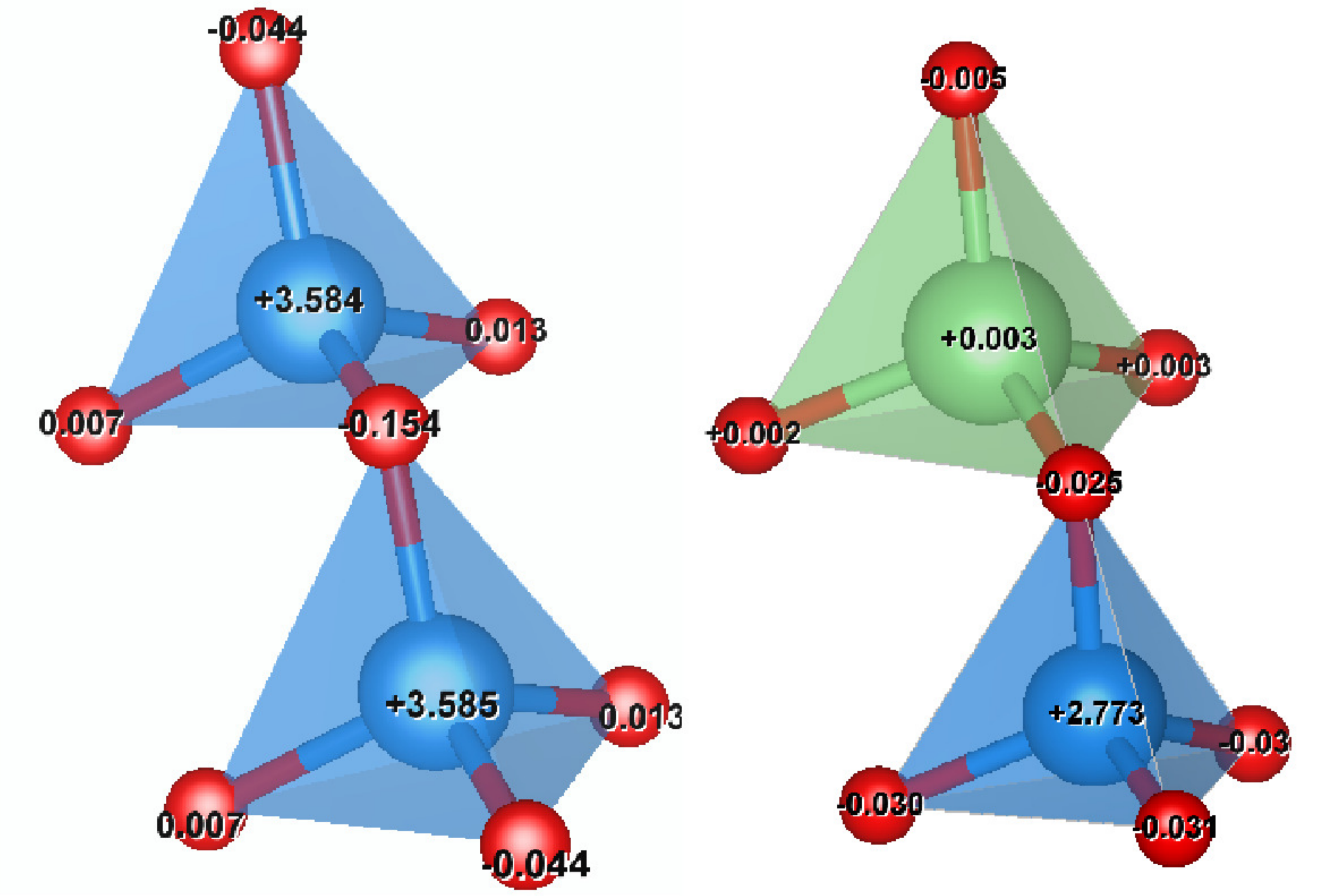}
\caption 
{Local magnetic moments at the Mn or Mg ions and the neighboring ligands for one pair of Mn-Mn ions (left) 
and one pair of Mn-Mg ions (right).}
\end{figure}

\begin{figure}
\centering%
\includegraphics[width=0.9\linewidth]{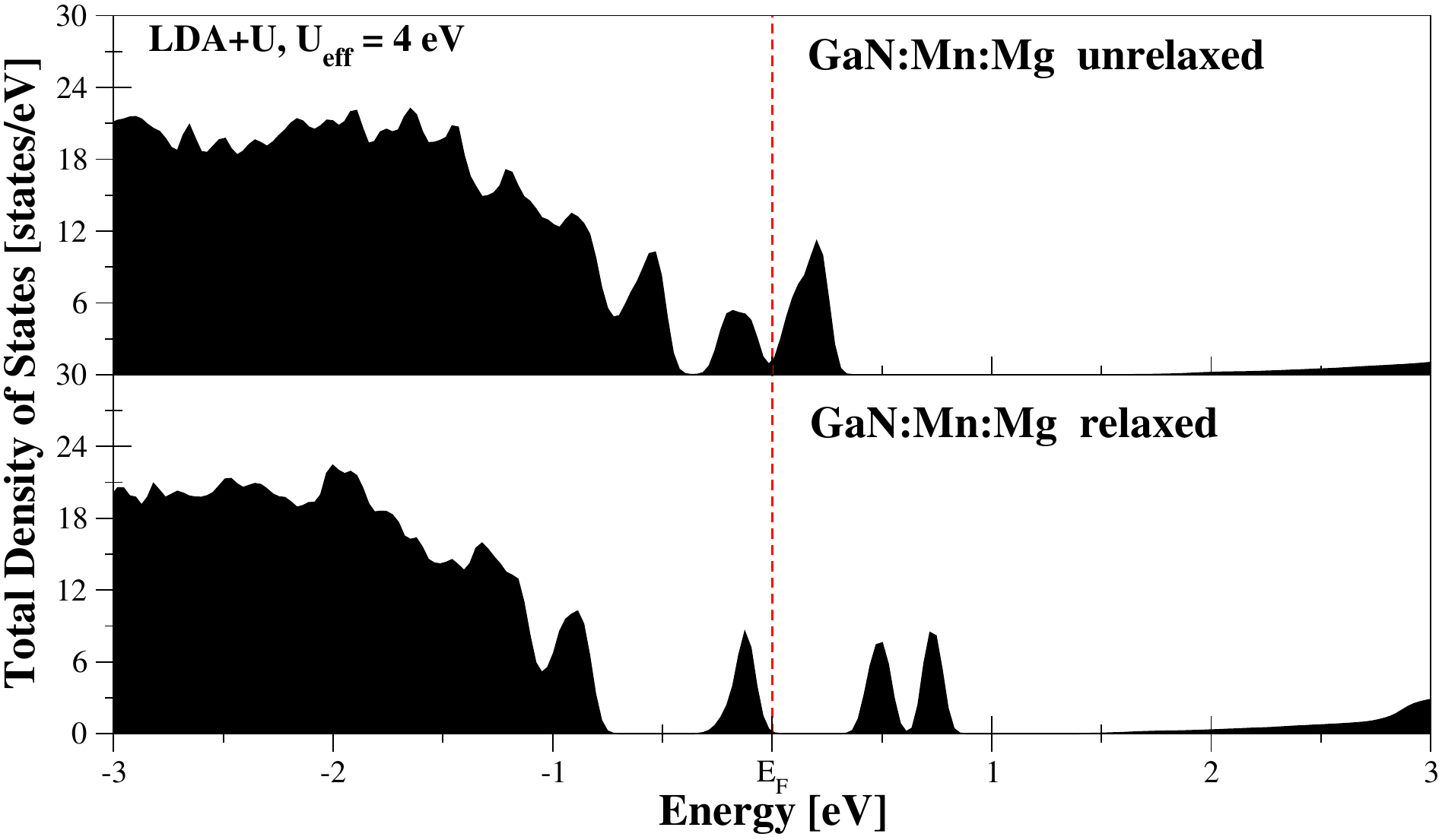}
\caption
{Total DOS projected onto one spin direction for the 64 atom supercell of GaN co-doped with 1 Mn and 1 Mg
at neighboring positions in the unrelaxed (top) and relaxed (bottom) structure.}
\end{figure}

\begin{figure}
\centering%
\includegraphics[width=0.9\linewidth]{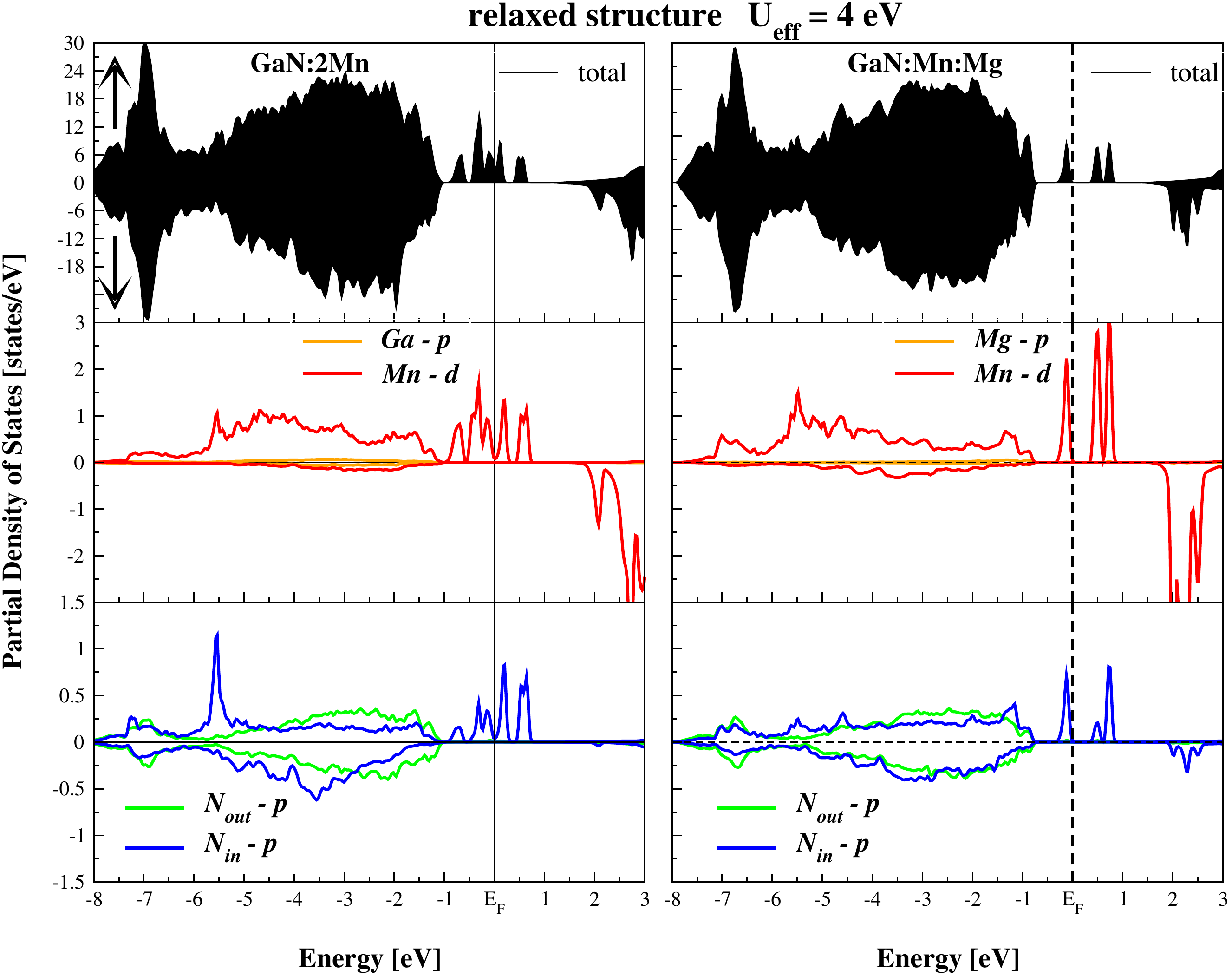}
\caption
{Total and partial DOS projected onto spin up (upper part) and spin down (lower part) for 
the relaxed structures of GaN with 2 Mn (left) or 1 Mn and 1 Mg (right). Shown are the partial DOS of the 
Mn $d$ orbitals, the Ga and Mg $p$ orbitals, as well as the N $p$ orbitals for a  
N atom close to the Mn impurity (N$_{in}$) and another one far of it  (N$_{out}$).} 
\end{figure}

\begin{figure}
\centering%
\includegraphics[width=0.9 \linewidth]{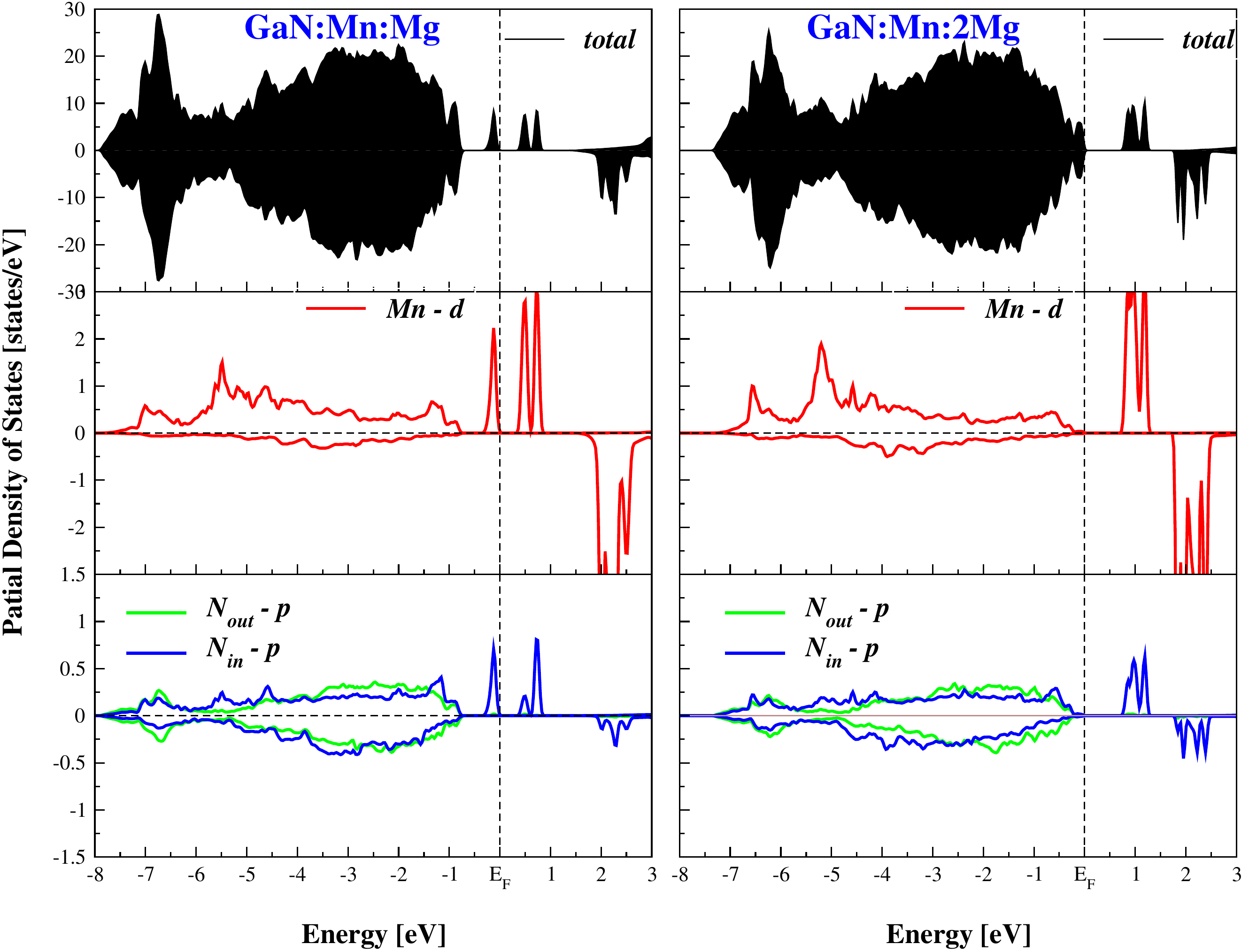}
\caption
{Same as Fig. 5 but comparing 1 Mn co-doped with 1 Mg (left) and 1 Mn co-doped
with 2 Mg (right).}
\end{figure}


\begin{table}
\begin{center}
\begin{tabular}{ | c | c | c  | c | c | } 
\hline
\hline
  \;   $U_{eff}$  \;   &  \; $M_{total} $ \; & \; \; $M_{Mn}$ \; \; & \; \; $M_{Mg} $ \; \; & \; \; $ M_{N} $ \; \; \\
(eV) & ($\mu_B$) & ($\mu_B$) & ($\mu_B$) &  ($\mu_B$) \\
\hline
0 & 3.007 & 2.289 & 0.004 & 0.035 \\
\hline
2 & 2.996 & 2.512 & 0.003 & 0.008 \\
\hline
4 & 3.097 & 2.773 & 0.003 & -0.025 \\
\hline
6 & 3.003 & 3.235 & 0.000 & -0.143 \\
\hline
8 & 3.006 & 3.657 & -0.002 & -0.219 \\
\hline
10 & 2.968 & 3.663 & -0.002 & -0.231 \\
\hline
\end{tabular}
\caption
{Magnetic moments for a complex consisting of 1 Mn and 1 Mg in GaN
as a function of $U_{eff}$.
Given are the total magnetic moment $M_{total}$ as well as the local 
moments at the Mn, Mg and the bridging N sites.}
\label{label}
\end{center}
\end{table}

\begin{table}
\begin{center}
\begin{tabular}{ | c | c | c  | c | c | } 
\hline
\hline
  \;   $U_{eff}$  \;   &  \; $M_{total} $ \; & \; \; $M_{Mn}$ \; \; & \; \; $M_{Mg} $ \; \; & \; \; $ M_{N} $ \; \; \\
(eV) & ($\mu_B$) & ($\mu_B$) & ($\mu_B$) &  ($\mu_B$) \\
\hline
4 & 2.039 & 2.124 & 0.002 & -0.050 \\
\hline
\end{tabular}
\caption
{Magnetic moments for a complex consisting of 1 Mn and 2 Mg in GaN
for $U_{eff}=4$ eV.
Given are the total magnetic moment $M_{total}$ as well as the local 
moments at the Mn, Mg and the bridging N sites.}
\label{label}
\end{center}
\end{table}

\begin{figure}
\centering%
\includegraphics[width=0.5\linewidth]{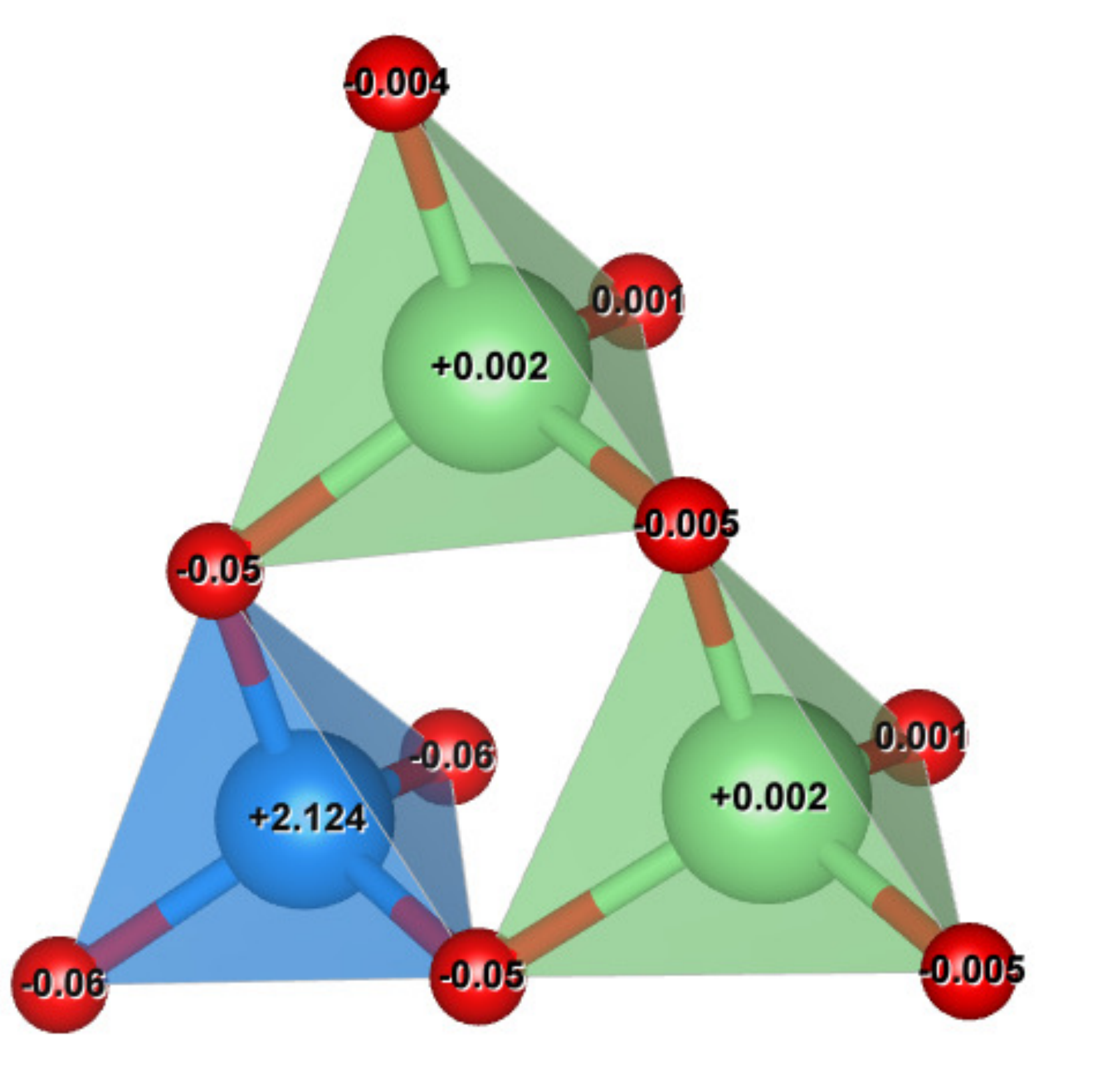}
\caption
{Local magnetic moments at the Mn, Mg ions and the neighboring ligands
for a complex of 1 Mn and 2 Mg.}
\end{figure}



\section{Co-doping with Magnesium}

To investigate the influence of co-doping Mn and Mg to GaN we studied a 64 atoms supercell of GaN with 1 Mn and
1 or 2 Mg ions. The corresponding total and partial DOS are shown in Figs. 5, 6, and 7. Fig. 5 shows nicely 
how a gap opens
due to relaxation in the supercell with 1 Mn and one Mg. Figs. 6 and 7 show the changes of total and partial DOS
by adding 1 Mg or 2 Mg, respectively. The corresponding magnetic moment distributions are presented in Tables 
III and IV.  All these results show that
co-doping of GaN with Mn and Mg leads to a valence change of Mn. Without Mg, the Mn valence is +3. That 
means that no charge carriers are doped into GaN. But by substituting one, or two Mg atoms for Ga, the 
Mn valence changes to +4 or +5, correspondingly. At the same time the local magnetic moment of Mn 
diminishes from a value corresponding to roughly $S=2$ (four unpaired electrons) to $S=3/2$ or 
$S=1$. 
That is best visible in the total magnetic moment values per Mn ion for $U=4$ eV 
(see Tables II, III, and IV) which decrease from 4 $\mu_B$ to 3.01 $\mu_B$ and 2.04 $\mu_B$
for the supercells without, with 1 or 2 Mg ions. 
The corresponding values of magnetization values
at the Mn sites are different, namely 3.58 $\mu_B$,  2.77 $\mu_B$ and 2.12 $\mu_B$, 
since part of the total magnetic moment is distributed on the neighboring N-sites with different orientation.
That is shown in Fig. 4 for the example of a pair of Mn or Mn/Mg impurities and also presented in Tables II-IV.

The magnetization values of Tables III and IV can be compared with the former study of 
Devillers of all. \cite{Devillers12} which
used the Quantum Espresso code. \cite{Giannozzi09} The general tendency is the same but 
the local values can be quite different.
For instance, the local Mn moment in the study of Villars et al for the complex with one Mn and one 
Mg is 4.02 $\mu_B$
and considerably larger than our values in Table III. That discrepancy can be explained by the different ways of 
calculating local moments in the Wien2\textit{k} or the Quantum Espresso code. 

Much more convincing, and free of methodological details is the analysis of the Mn valency by means of the DOS
(see Figs. 6 and 7). 
The Mn valency is determined by the filling of the above mentioned $t_{2g}$ 
peaks at Fermi level.
As it was already mentioned, the filling of the $t_{2g}$ peaks with two electrons corresponds
to a Mn valency of 3+. If we now dope additionally one Mg ion, the $t_{2g}$ multiplet 
can only be filled with one electron. And since it 
is the Mn 3$d$ peak which is located at the Fermi level and since that peak is very well separated from the valence
band even for $U$ values between 2 and 8 eV (in difference to the situation in GaAs:Mn) it is quite 
natural that the doped
holes due to Mg doping sit at Mn and are not distributed in the valence band. That is already visible in the DOS.
Correspondingly, the Mn valency is 4+ for the supercell with one Mg. And it is a 
logic consequence, that doping a second
Mg ion into the supercell leads to the loss of the last electron in the $t_{2g}$ level and to the Mn$^{5+}$ 
valence.


\section{Conclusion}

We analyzed numerically a 64 atom supercell of zinc-blende GaN with 1 or 2 Mn ions, as well as with 1 Mn ion
and 1 or 2 Mg ions. The nearest neighbor exchange of two Mn was found to be ferromagnetic with the magnetization 
density concentrated in between the two Mn sites (see Figure 3). The exchange coupling constant $J$ 
(in the Hamiltonian $\hat{H}=-2 J \hat{\vec{S}}_1\cdot \hat{\vec{S}}_2$ )
was found to be about 18 meV. The lattice relaxation leads to a decrease of 
the density of states at the Fermi level (pseudo gap feature, Fig.\ 2) but does not influence the 
magnetic coupling in a significant way.

Co-doping with Mg leads to holes which sit, however, at the Mn sites and change the Mn valency from 
Mn$^{3+}$
without Mg to Mn$^{4+}$ with 1 Mg and Mn$^{5+}$ with 2 
Mg. We showed that this valence change can already be concluded 
from a careful analysis of the DOS without Mg co-doping since the relevant peak at Fermi level is of Mn character,
well separated from valence or conduction band and filled with 2 electrons for GaN:Mn.
 
We can even deduce more conclusions from our analysis of the DOS. Let us suppose a 
complex with 1 Mn and 3 Mg ions in GaN 
which would lead to further $p$ doping. But the Fermi level is then expected to enter into the main valence band
with low Mn character. As a consequence, the valence of Mn is expected to stay 5+ for such a complex 
which explains
the experimental observation. \cite{Devillers12} It should also be noted that the Mn-Mg complexes in GaN 
show highly
interesting optical properties especially in the near infrared. Even stimulated emission could recently be 
shown.\cite{Capuzzo17} The 
analysis of the optical properties of these materials could be an interesting subject of further studies.

\section{Acknowledgements}

We thank NATO SPS program (grant SfP-984735) for financial support. R.H. thanks A. 
Bonanni, D. Kysylychyn, A. Nikolenko, and V. Strelchuk  for 
stimulating discussions.

\end{document}